\documentclass[12pt]{spieman}  
\usepackage{amsmath,amsfonts,amssymb}
\usepackage{graphicx}
\usepackage{setspace}
\usepackage{tocloft}

\title{Enhanced Light-Matter Interactions in Dielectric Nanostructures via Machine Learning Approach}

\author[a]{Lei Xu}
\author[b,*]{Mohsen Rahmani}
\author[a]{Yixuan Ma}
\author[b]{Daria A. Smirnova}
\author[b]{Khosro Zangeneh Kamali}
\author[a]{Fu Deng}
\author[a]{Yan Kei Chiang}
\author[a]{Lujun Huang}
\author[c]{Haoyang Zhang}
\author[d]{Stephen Gould}
\author[b]{Dragomir N. Neshev}
\author[a,*]{Andrey E. Miroshnichenko}
\affil[a]{School of Engineering and Information Technology, University of New South Wales, Canberra, ACT 2600, Australia}
\affil[b]{Nonlinear Physics Centre, Research School of Physics, The Australian National University, Canberra, ACT 2601, Australia}
\affil[c]{School of Electrical Engineering and Computer Science, The Queensland University of Technology, Brisbane, QLD 4000, Australia}
\affil[d]{ANU College of Engineering and Computer Science, The Australian National University, Canberra,  ACT  2601, Australia}

\cftpagenumbersoff{figure}
\cftpagenumbersoff{table} 
\begin{document} 
\maketitle

\begin{abstract}
A key concept underlying the specific functionalities of metasurfaces is the use of constituent components to shape the wavefront of the light, on-demand.  Metasurfaces are versatile and novel platforms to manipulate the scattering, colour, phase or the intensity of the light. Currently, one of the typical approaches for designing a metasurface is to optimize one or two variables, among a vast number of fixed parameters, such as various materials’ properties and coupling effects, as well as the geometrical parameters. Ideally, it would require a multi-dimensional space optimization through direct numerical simulations. Recently, an alternative approach became quite popular, allowing to reduce the computational cost significantly based on a deep-learning-assisted method. In this paper, we utilize a deep-learning approach for obtaining high-quality factor (high-Q) resonances with desired characteristics, such as linewidth, amplitude and spectral position. We exploit such high-Q resonances for the enhanced light-matter interaction in nonlinear optical metasurfaces and optomechanical vibrations, simultaneously. We demonstrate that optimized metasurfaces lead up to 400+ folds enhancement of the third harmonic generation (THG); at the same time, they also contribute to 100+ folds enhancement in optomechanical vibrations. This approach can be further used to realize structures with unconventional scattering responses.  
\end{abstract}

\keywords{Machine learning, Dielectric nanostructures, Fano resonance, Third harmonic generation, Opto-acoustic.}

{\noindent \footnotesize\textbf{*}Mohsen Rahmani, \linkable{mohsen.rahmani@anu.edu.au}; Andrey E. Miroshnichenko,  \linkable{andrey.miroshnichenko@unsw.edu.au} }

\begin{spacing}{2}   

\section{Introduction}
\label{sect:intro}  
Metasurfaces are thin and flat arrays of subwavelength nanoparticles, enabling control over the polarization, phase, amplitude, and dispersion of light~\cite{Neshev2018Optical}. They can be used for light emission, detection, modulation, control and/or amplification at the nanoscale. In recent years, metasurfaces have been a subject of undergoing intense study as their optical properties can be adapted to a diverse set of applications, including superlenses, tunable images, holograms, etc~\cite{Neshev2018Optical,hail2019optical,chang2018optical,Kuznetsov2016Optically,rahmani2017reversible}. High-refractive-index dielectric metasurfaces provide a powerful platform for controlling light that can go beyond plasmonics, as they cause negligible losses as compared with plasmonic metasurfaces. Dielectric materials offer unique ability to efficiently manipulate light at the nanoscale based on the simultaneous excitation and control over the optically induced electric and magnetic Mie-type resonances. 
Resonant dielectric metasurfaces with high-quality factor (high-Q) resonances, in particular, are of significant interests due to their possibilities to strongly enhance the electromagnetic near-fields and boost the light-matter interactions at the nanoscale. In other words, they allow to enhance the response of metasurfaces to an external electromagnetic field at a particular frequency. Moreover, high-Q metasurfaces can increase the storage time of photons and, thus, light-matter interactions within the subwavelength resonators. It will facilitate various nanophotonics applications, such as enhanced nonlinear photon generations, optical sensing, optoacoustic vibrations as well as narrowband filtering. In the last decade, high-Q metasurfaces were mainly associated with Fano resonances (FRs) featuring asymmetric spectral line profiles~\cite{miroshnichenko2010fano,luk2010fano,khardikov2012giant,rahmani2013fano,gupta2016toroidal}. In FRs, the asymmetry originates from a close interaction of a discrete (localized) state with a continuum of propagation modes~\cite{Yang2015Nonlinear,tuz2018high,campione2016broken}. Some examples are trapped modes in arrays of dielectric nanodisks with asymmetric holes~\cite{zangeneh2019reversible,xu2019dynamic,jain2015electric}, ring and disk cavities~\cite{hao2009tunability}, Dolmen structures~\cite{liu2011three,hao2008symmetry}, and aggregated nanoparticles~\cite{bao2010fano,halas2011plasmons,hentschel2010transition,rahmani2012subgroup}. Recently, it has been shown that different approaches for generating high-Q resonances are based on the bound-state-in-the-continuum (BIC), i.e., a localized state with zero linewidth that is embedded in the continuum\textcolor[rgb]{1,0,0}{~\cite{hsu2016bound,carletti2018giant,koshelev2018asymmetric,liu2018extreme,he2018toroidal,xu2019dynamic,Koshelev2019,luo2020excitation,carletti2019high,gandhi2020gain}}. Indeed, optical BICs provide a unique opportunity to manipulate the light-matter interaction within the radiative continuum because of their ultrahigh-Q origin and associated giant enhancement of the electromagnetic near-field~\cite{ndangali2010electromagnetic}. 

On the other hand, designing metasurfaces with high-Q resonances are usually achieved via continuous parameters tuning, with limited control on the linewidth, amplitude, and spectral positions. Currently, one of the typical approaches for designing metasurfaces with desired resonance is based on direct optimization of one or two parameters via brute-force simulations~\cite{Gili2016Monolithic,Liu2016Resonantly,Camacho-Morales2016Nonlinear,Grinblat2016Efficient,Grinblat2016Enhanced,shcherbakov2015ultrafast,Melik-Gaykazyan2017Third,melik2017selective,miroshnichenko2015nonradiating,Xu2018Boosting,Carletti2015Enhanced}. This is a time-consuming task accompanied by a random success on the output parameters of the desired resonances~\cite{molesky2018inverse}. Recently, deep learning approaches, based on the artificial neural networks (ANNs), have emerged as a revolutionary and robust methodology in nanophotonics~\cite{liu2018training,mirzaei2014superscattering,nadell2019deep,kiarashinejad2020deep,ma2018deep,yao2019intelligent,liu2019integrated,gao2019bidirectional,zhang2019artificial,jiang2019simulator,so2019simultaneous,jiang2019global,wiecha2019pushing,piggott2015inverse,asano2018optimization}. Indeed, applying the deep learning algorithms to the nanophotonic inverse design can introduce remarkable design flexibility that can go far beyond that of the conventional methods\textcolor[rgb]{1,0,0}{~\cite{carleo2019machine}}. The inverse design approach works based one the training process, that enables fast prediction of complex optical properties of nanostructures with intricate architectures. 

As a (non-unique) example, we have targeted toroidal dipoles, due to their promising applications in the formation of anapole states and electromagnetic energy localization. We investigate the non-radiating toroidal dipole (TD) supported by two parallel silicon bars, as the building blocks of the metasurface. The reason behind choosing such a geometry is the reasonable number of parameters to be optimized, as a proof of concept. The parameters include the length and width of the bars, as well as the gap between them. At the same time, this TD corresponds to the symmetry-protected BIC. Subsequently, via taking this TD-BIC model, we demonstrate the deep-learning-assisted inverse design of arbitrary high-Q resonances with different line width, amplitude and spectral position. We employ a multi-layer perceptron (MLP) variant of the ANN as our model~\cite{carleo2017solving,hertz2018introduction,hornik1989multilayer}. An MLP consists of multiple layers of perceptrons (MLP), including an input and an output layer with several hidden layers. In this work, each artificial neuron in one layer connects with a certain weight to every neuron in the following layer, that are adapted during the learning state. Once learned, the weight values remain fixed, and the model can be used to infer the target metasurface design parameters. The hidden layers establish a nonlinear mapping between the input and output through training from the given dataset, and then be able to predict the response of the system, or inversely determine the design parameters for the desired performance.

The proposed deep-learning-assisted inverse-design approach provides a platform to design metasurfaces for more than one application. In this paper, we employ machine learning to obtain a bi-functional metasurface dealing with photons and phonons, simultaneously~\cite{Aouani2014Third,lee2014giant,merklein2015enhancing,pant2011chip,VanLaer2015,Grinblat2014,DellaPicca2016,Aouani2015,Chen2011,Moreaux2001}. Photon-photon conversions, so-called nonlinear nanophotonics is at the heart of modern macroscopic optics, including lasers, sensors, imaging and information technology. On the other hand, photon-phonon conversions are the state-of-the-art solution for precision mass sensors, micro-manipulation, and sensing biochemical materials, with transformative implications in the fields of health and security. A combined photon-phonon conversion can be used for non-ionizing and non-invasive imaging~\cite{Chen2011}. Here, by using the deep-learning-assisted inverse approach we design and fabricate a single optoacoustic metasurface that enhances third-harmonic generation intensity for 400+ folds and acoustic mode excitations for 100+ folds, concurrently, all through a designed high-Q resonance, associated with a strong electric near-field enhancement. The inverse design approach proposed in this paper is extendable to other characteristics and applications of metasurfaces and significantly circumvents the time-consuming, case-by-case numerical models in conventional electromagnetic nanostructure designs.

\section{Results and Discussions}

\begin{figure}[h!]
\centering
\includegraphics[width=0.8\textwidth]{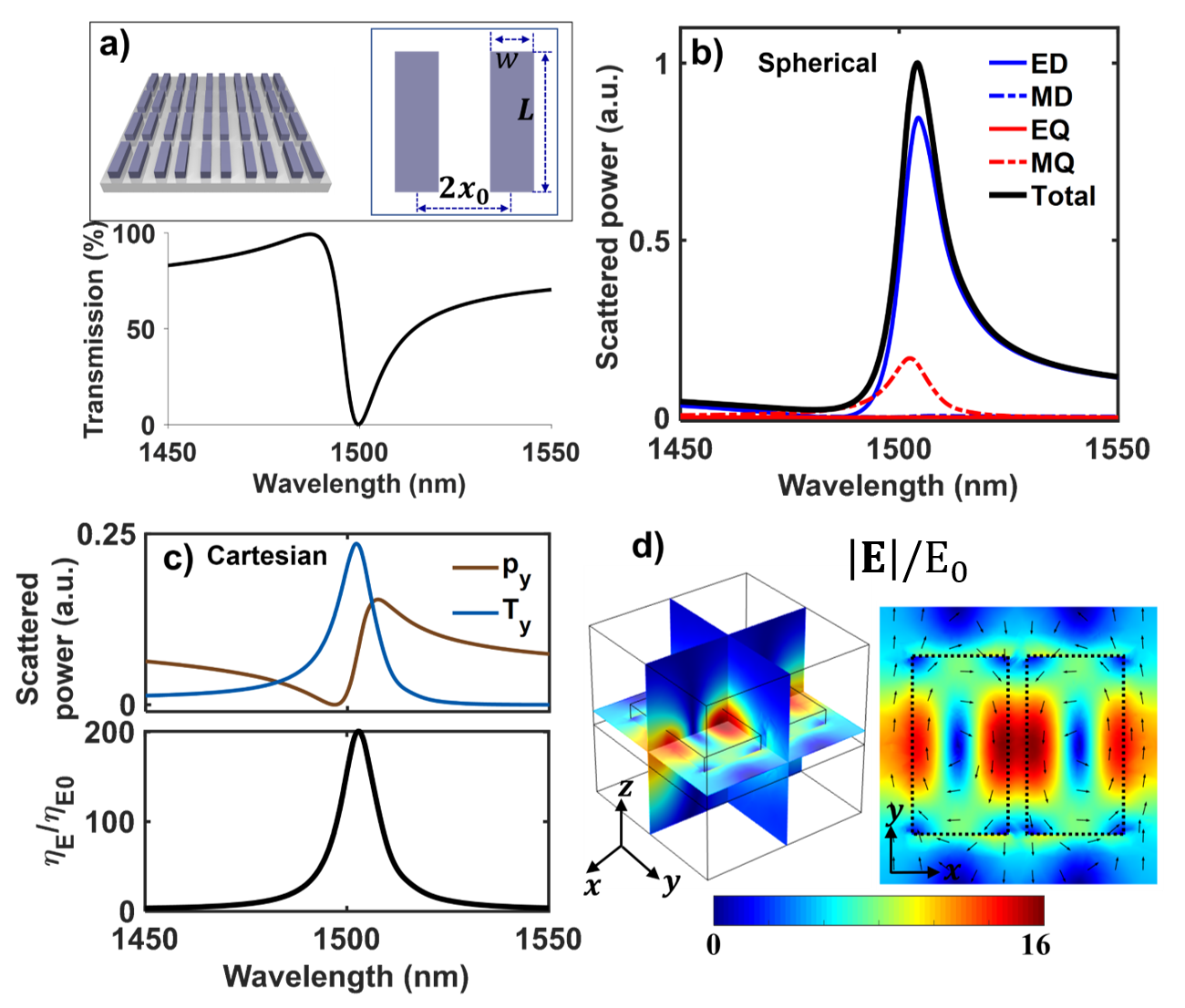}
\caption{(a) Top: Schematics of the silicon nanobars metasurface (top left), and its unit cell (top right); Bottom: Calculated transmission spectrum of the metasurface with structural parameters $w=316$~nm, $L=580$~nm, $x_0=189$~nm.  (b) Spherical multipolar structure of the metasurface. (c) Top: Cartesian electric dipole and toroidal dipole modes excitations. Bottom: The electric energy enhancement $\eta_E/\eta_{E_{0}}$. It is defined as the electric energy inside the two nanobars normalized by the electric energy within the same volume of the nanobars for pump field. (d) Electric near-field distributions at the resonance. Left: 3-dimensional view. Right: top view.}
\label{fig:1}
\end{figure}

To obtain the initial high-Q resonances, we have defined the building blocks of metasurfaces to be two identical silicon nanobars with width $w$, length $L$ and the offset $x_0$, which is the distance between the center of the two bars fabricated on a glass substrate, as shown in the top panel of Fig.~\ref{fig:1}(a). It is worth mentioning that there is a large variety of other geometries, demonstrated earlier for generating high-Q resonances~\cite{shcherbakov2015ultrafast,Liu2016Resonantly,hsu2016bound,carletti2018giant,koshelev2018asymmetric,xu2019dynamic,liu2018extreme,he2018toroidal,Koshelev2019,luo2020excitation,carletti2019high}. However, discussions on advantages and disadvantages of various geometries are beyond the scope of the current study. We rather concentrate on customizing of the generated high-Q resonances.  The thickness of bars is fixed at 150~nm, and the periodicity of unit cells is fixed at $D=900$~nm. in both $x$ and $y$ directions. As can be seen in the bottom panel of Fig.~\ref{fig:1}(a), this structure supports a strong Fano resonance in the transmission spectrum, when the incident light is polarized along $y$-axis. This Fano resonance is formed by the interference and coupling between a ``bright'' electric dipole resonance $p_y$ and a ``dark'' toroidal dipole mode $T_y$. Fig.~\ref{fig:1}(b) shows the corresponding spherical multipolar decomposition of the metasurface. As can be seen, the optical response is dominated by electric dipole (ED) excitation with a small contribution from the magnetic quadrupole (MQ) resonance. Such a pronounced ED feature was further investigated by performing the Cartesian multipolar analysis (Fig.~\ref{fig:1}(c)). It is worth noting that the ED response is mainly due to the strong excitation of toroidal dipole mode $T_y$ with an in-plane ED mode $p_y$, which is polarized along the same direction as the optical pump. 

Interestingly, due to the $C_2$ symmetry of our sub-diffractive system, the toroidal dipole (TD) and MQ do not contribute to the far-field radiation along the $z$-direction. The far-field optical response is dominated by the ED mode $p_y$. The non-radiating TD mode is a symmetry-protected BIC, where the ED mode $p_y$ plays a role to open a leaky channel and transform this ideal BIC into quasi-BIC with a finite Q-factor. 
In our previous work~\cite{xu2019dynamic}, we studied the formation of BIC enabled by a magnetic dipole (MD) resonance in Si disk-with-hole metasurfaces, where geometrical asymmetry was introduced to open a leaky channel.
In contrast, here we show that the leaky channel, i.e., the excitation of ED mode $p_y$, can be formed directly by properly choosing the structural dimensions of the symmetric nanobars. Owing to the non-radiating nature of the dominant resonance – TD mode, a clear enhancement of the stored electric energy inside the nanobars is observed, as shown in the bottom of Fig.~\ref{fig:1}(c). Fig.~\ref{fig:1}(d) gives the calculated electric near-field distributions. A pronounced poloidal current distribution can be observed from the two nanobars, indicating strong TD excitation. The small portion of MQ excitation shown in Fig.~\ref{fig:1}(d), is due to the uncompensated circulating magnetic field in such flat geometry, formed by two anti-parallel magnetic dipole moments at the nodes of the poloidal current distribution. A comparison between the electric near-fields between the y-polarized pump and $x$-polarized pump incidence can be seen in Fig.~S1 of the Supporting Information. We observed significant near-field enhancement inside the nanostructure for y-polarized pump incidence as compared to the case for $x$-polarized pump incidence. A further investigation on the band structure and the corresponding mode profiles can be seen in Fig.~S2 and~S3 of Section I in the Supporting Information.

As mentioned earlier, achieving scalable metasurfaces with several interdependent characteristics, including quality factor and spectral position, is the main target of this manuscript. There is a significant demand in the photonics community to achieve both conditions, rather than solely obtaining a high-Q resonance. In this respect, we employ the deep learning approach to inversely design our metasurfaces with simultaneously controlling and optimizing on Q-factor, amplitude and spectral position. For this task, we use the open-source neural-network library Keras~\cite{Chollet2015} written in Python to implement our method.

\begin{figure}[h!]
\centering
\includegraphics[width=0.75\textwidth]{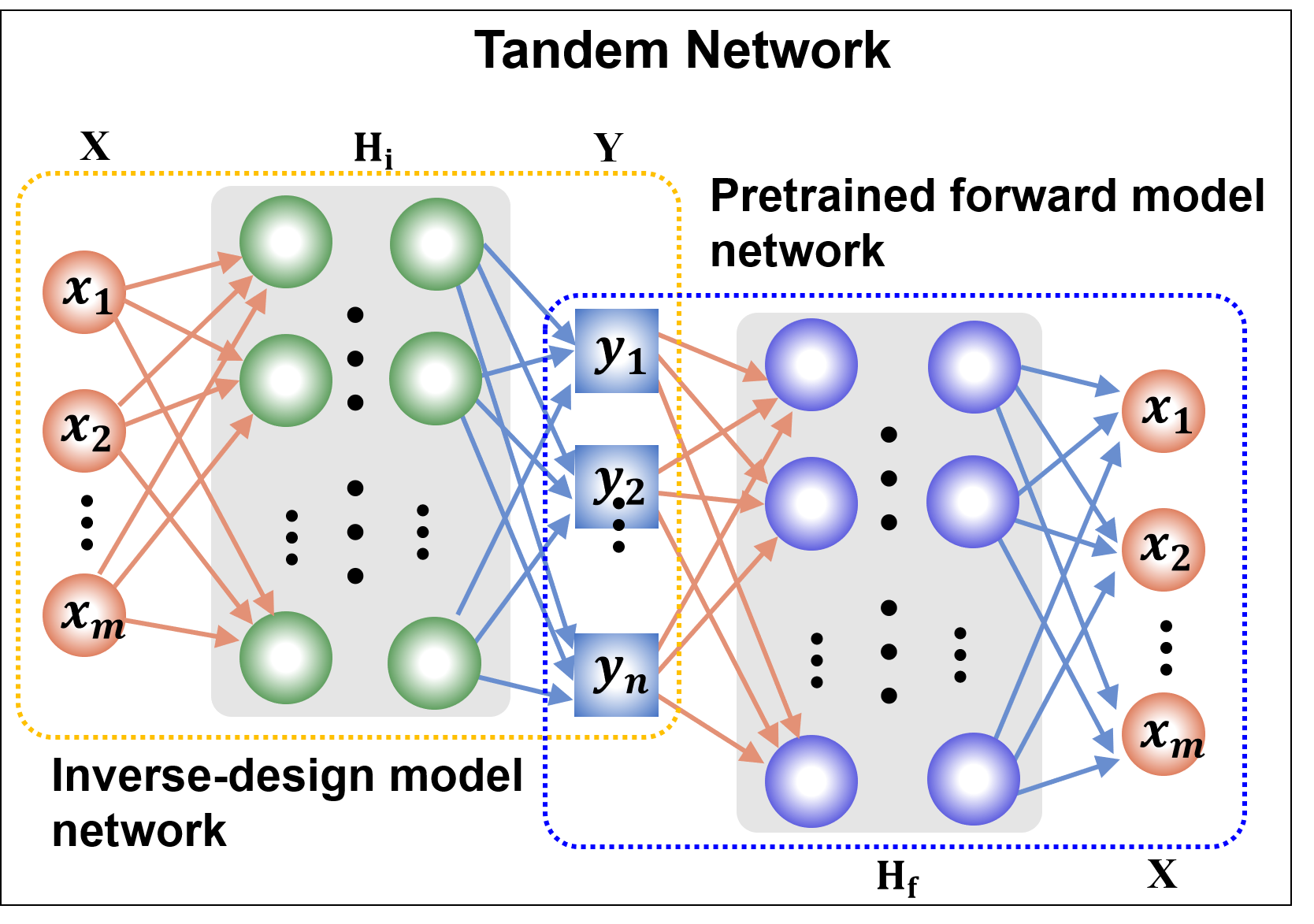}
\caption{The architecture of the Tandem Network model which consists of an inverse design network connected to a pre-trained forward model network. $X$ represents the input and output, which is the transmission spectra data in our case, and $Y$ represent the output in the middle layer which is the structural parameters here.}
\label{fig:2}
\end{figure}

For the inverse design of nanophotonic structures using a deep learning approach, one challenge is that the same far-field electromagnetic response can correspond to different designs, i.e., several different structures can give the same responses. This nonuniqueness of the response-to-design mapping will induce conflicting examples within the training set and may prevent from converging. To avoid this issue, we use the Tandem Network (TN) approach~\cite{liu2018training,so2019simultaneous}, as shown in Fig.~\ref{fig:2}. The TN architecture consists of an inverse design network connected to a forward model network. The forward network learns the mapping from the structural parameters to the optical responses and is trained separately first. After the forward network is trained, it is placed after the inverse-design model network, and its network weights keep fixed during the training of the inverse-design model network. The inverse-design network learns a mapping from the optical responses to the structural parameters. When training the inverse-design network, its weights are updated to minimize the loss objective: $J=\frac{1}{n}\sum_i(r_i-o_i)^2$  with $r_i$ and $o_i$ being the values predicted by the neural network and the ground truth of the response (forward model network) or the structural parameters (inverse-design model network). Trained in this way, the inverse-design network is not constrained to produce a pre-specified design. Instead it is free to infer any design that results in the desired forward behaviour.

In our case, for simplicity, we specify three structural parameters of the nanobars to learn: width $w$, length $L$ and the offset distance $x_0$. It is worth mentioning that machine learning approach can be extended to predict nanostructures with more parameters or even different types of parameters such as materials properties and materials losses, etc.
By randomly specifying them, we first use the rigorous coupled-wave analysis (RCWA)~\cite{Moharam1995} to generate 25,000 training examples, where we obtain the transmission spectra of the metasurfaces covering wavelength range 1400~nm to 1600~nm. Here, as the refractive index of silicon in this wavelength range is nearly constant, thus in our simulation, we keep the refractive index of silicon as 3.6 to expedite the training data generation process. RCWA is a frequency-domain modal method based on the decomposition of the periodic structure and the pseudo-periodic solution of Maxwell's equations in terms of their Fourier expansions~\cite{Hugonin2005rcwa}. It has been widely used for modelling light responses from periodic optical structures due to its fast convergence and accurate far-field calculations~\cite{Arbabi2015}. It is quite suitable for modelling the electromagnetic responses of metasurfaces and generates massive training data, especially when considering the inverse design, based on neural networks~\cite{liu2018training,yao2019intelligent}. It is worth mentioning that for multilayer structures, the transfer matrix method can be one more choice~\cite{Peurifoy2018}.

\begin{figure}[h!]
\centering
\includegraphics[width=0.8\textwidth]{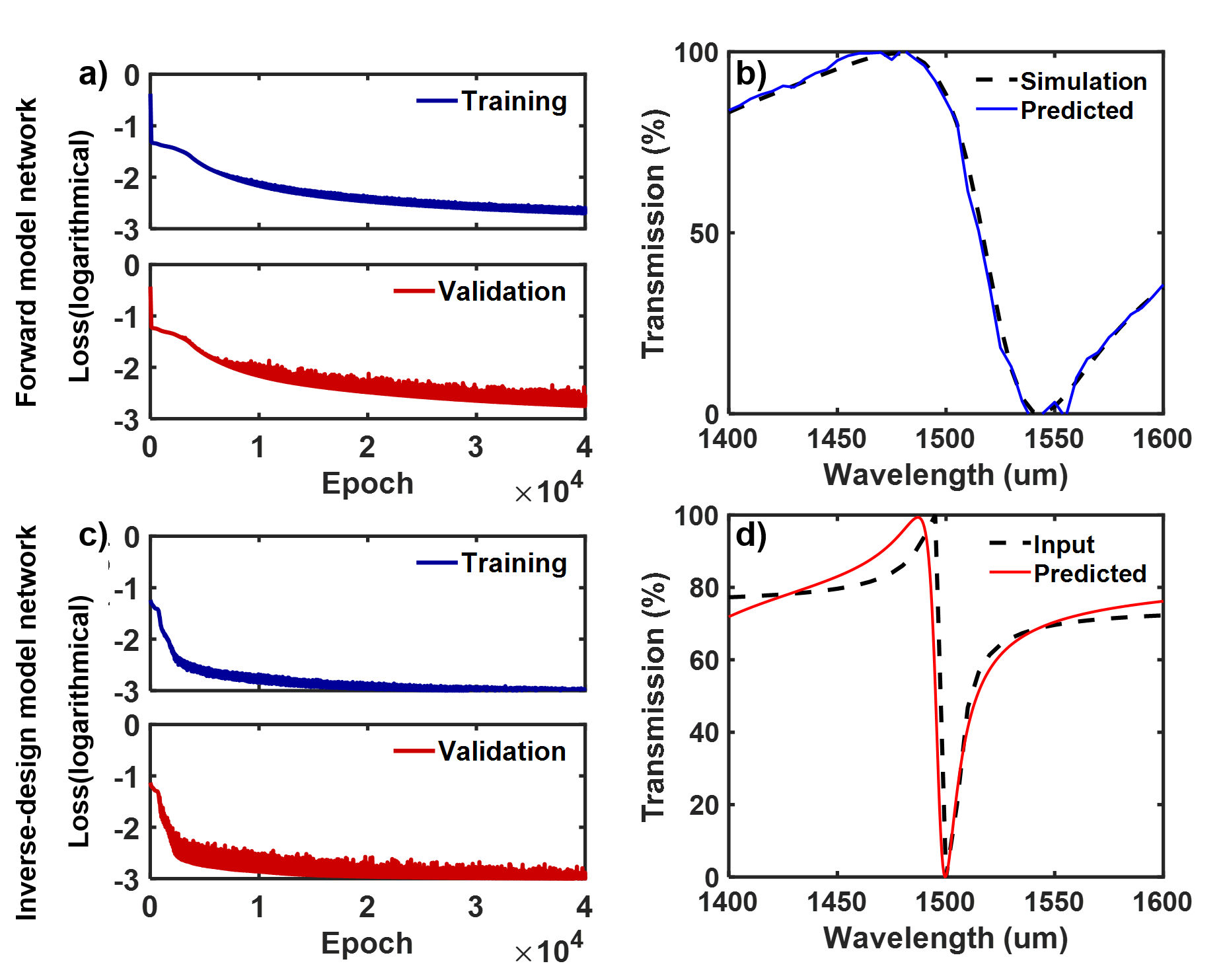}
\caption{(a) Evolution of the training loss for the forward model network. (b) Comparison of the NN approximation to the real transmission spectrum. (c) Evolution of the training loss for the inverse-design model network. (d) Comparison of the spectra between the NN approximation and the input based on Eq.~\ref{eq:Fano2}.}
\label{fig:NN1}
\end{figure}

The forward model network is designed to have four fully-connected layers with each layer having 400-600-400-200 dimensions, respectively.  We set the learning parameters batch size as 256, and use a learning rate of 0.001 and decay of $1\times10^{-6}$. We first train the forward model network and evaluate it to see how well it can predict the given transmission spectra. Fig.~\ref{fig:NN1}(a) shows the learning curves for training and validation loss. It can be seen that both the training loss and validation loss decrease significantly after 10,000 epochs of training and become less than 0.005 after 30,000 epochs of training. It indicates that the trained network can estimate an appropriate spectrum, which is similar to the spectrum calculated analytically. As a test example, we use RCWA to simulate the transmission spectrum for an individual Si nanobar metasurface with parameters $[w, L, x_0]$ being [300, 700, 300]~nm, as shown by the black dashed curve of Fig.~\ref{fig:NN1}(b). Then, we input these structural parameters to the network and predict the corresponding output, which is shown in the blue curve of Fig.~\ref{fig:NN1}(b). As can be seen, our forward network can predict the transmission spectrum from our metasurface accurately. We have also tested our forward network by specifying different structural parameter sets corresponding to different transmission spectra (see Fig.~S4 in the Supporting Information), which all verify the effectiveness of our forward network.

Next, we train the inverse-design model network by fixing the weights in the pre-trained forward model network. Since the forward model network is differentiable, we can train the inverse-design model network with a loss placed after the forward model network. As mentioned above, this will overcome the issue of non-uniqueness in the inverse spectrum of electromagnetic waves, as the design by the neural network is not required to be identical to the design parameters that produced the training samples, only that the spectrum inferred by the forward model network match the target spectrum. The loss can be further lowered when the generated design and the real design have similar responses after training. The inverse-design model network has five fully-connected layers with 600, 600, 400, 200, 200 dimensions, respectively. For the inverse design of our metasurfaces, the transmission spectrum is considered as the input of the tandem network. The design parameters are predicted from the intermediate layer of the whole network. The training process is shown in Fig.~\ref{fig:NN1}(c). As can be seen, by using TN approach, the learning of inverse design has converged effectively. We then test our inverse-design network by using a Fano formula to define the transmission spectrum~\cite{Maksymov2011,Galli2009}:
\begin{equation} \label{eq:Fano}
    F(\omega)=A_0+F_0\frac{\left[q+2(\omega-\omega_0)/\Gamma\right]^2}{1+\left[2(\omega-\omega_0)/\Gamma\right]^2}
\end{equation}
where $\omega_0$ and $\Gamma$ are the resonance frequency and linewidth, respectively. $A_0$ and $F_0$ are constant factors and fixed at 1 in the rest of this paper. $q$ is a dimensionless factor that describes the ratio between the resonant and non-resonant transition amplitudes in the spectrum.
Here, our target is a Fano resonance with a peak value of 100\% and dip value of 0\%. Therefore, the resonance frequency and linewidth of the transmission resonances can be obtained via:

\begin{equation} \label{eq:Fano2}
    T(\omega)=\frac{F(\omega)-F_{min}}{F_{max}-F_{min}}
\end{equation}

We first design a target Fano-resonance at $\lambda_0=1500$~nm, with resonance linewidth $\Delta\lambda$=5 nm, and $q$=0.5, as shown by the black dashed curve of Fig.~\ref{fig:NN1}(d). Here, we input the resonance frequency linewidth $\Gamma$ by the defined resonance wavelength linewidth $\Delta\lambda$ by $\Gamma=\Delta\lambda\omega_0/\lambda_0$. We then use the network to predict the structural parameters $[w, L, x_0]$ of the required metasurface as $[316, 580, 189]$~nm. The transmission spectrum of the predicted metasurface is shown in the dashed red curve of Fig.~\ref{fig:NN1}(d). It matches well with the desired Fano-shape curve based on Equations~\ref{eq:Fano} and~\ref{eq:Fano2}.

\begin{figure}[h!]
\centering
\includegraphics[width=0.85\textwidth]{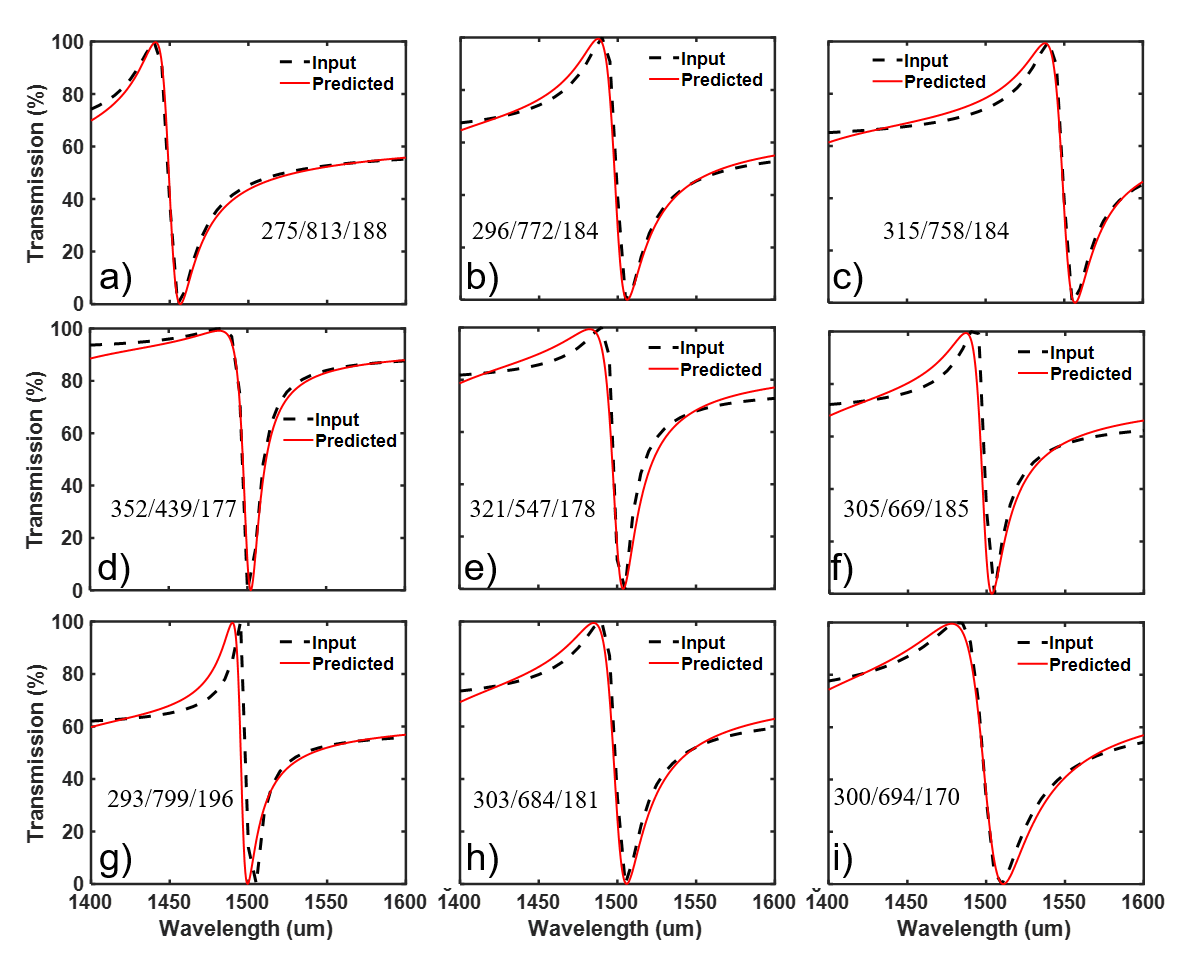}
\caption{Inverse design of Si nanobar metasurfaces with Fano-shape transmission spectra. (a-c) $\lambda_0=1450$~nm, 1500~nm and 1550~nm, respectively. $\Delta\lambda=15$~nm, $q=0.8$. (d-f) $\lambda_0=1500$ nm, $\Delta\lambda$ =10~nm, $q$=0.3, 0.5 and 0.7, respectively. (g-i) $\lambda_0$=1500 nm, $\Delta\lambda$ =5~nm, 15~nm and 25~nm, respectively, $q$=0.7.}
\label{fig:inversedesign}
\end{figure}

The deep network provides a powerful approach to design nanophotonics structures inversely. Once the training process is finished, the inverse design calculation takes only around 0.05 second in our case (only using CPU in a normal desktop with 64-bit Operating System: Processor: Intel(R) Core(TM) i7-4770 CPU @ 3.40 GHz, RAM: 16.0 GB), which is both promising and much more effective, as compared to conventional electromagnetic solvers. For nanophotonics applications, as mentioned above, it is a requirement to obtain scalable metasurfaces with controllable characteristics. In the following, using the TN approach, we further test the inverse design of Fano resonances with different spectral positions, linewidths or amplitudes.
According to Eq.~\ref{eq:Fano} and~\ref{eq:Fano2}, we first specify target Fano resonances at $\lambda_0$=1450~nm,1500~nm,and 1550~nm, respectively. Subsequently, we keep the linewidth $\Delta\lambda$=15~nm and $q$=0.8, as shown in the black dashed curve of Figs.~\ref{fig:inversedesign}(a-c). We then use the trained neural network to predict the structural parameters of the metasurfaces that can provide such spectra. The transmission spectra of the designed metasurfaces are shown in the red curves of Figs.~\ref{fig:inversedesign}(a)-\ref{fig:inversedesign}(c). They match and satisfy the design goal well. By varying the value of parameter $q$ or the linewidth of the resonance $\Delta\lambda$, the neural network can easily predict the metasurface design for the required Fano resonance with different characteristics, as shown in Figs.~\ref{fig:inversedesign}(d)-\ref{fig:inversedesign}(i). It provides a powerful method to control the near-field and electric energy confinement at the nanoscale (see Fig.~S5 in the Supporting Information).

So far, based on the deep learning approach, we have optimized the design to obtain high-Q resonances via altering various parameters, including the linewidth, spectral position and amplitude, simultaneously. Such a high-Q resonance can significantly facilitate nanostructures to enhance light-matter interactions for various applications such as nonlinear optics, optomechanics, etc. As an example, here, we investigate the third-harmonic generation (THG) from three designed metasurfaces with resonances at 1450~nm, 1500~nm and 1550~nm respectively, as shown in Fig.~\ref{fig:inversedesign}(a)-\ref{fig:inversedesign}(c). Here, we have taken into account that the spectral full-width-at-half-maximum (FWHM) of our experimentally used laser is around 15~nm for the wavelength range 1400 nm to 1600 nm. Thus, we consider the resonances with this linewidth to maximize the nonlinear signal generation.

\begin{figure}[h!]
\centering
\includegraphics[width=0.6\textwidth]{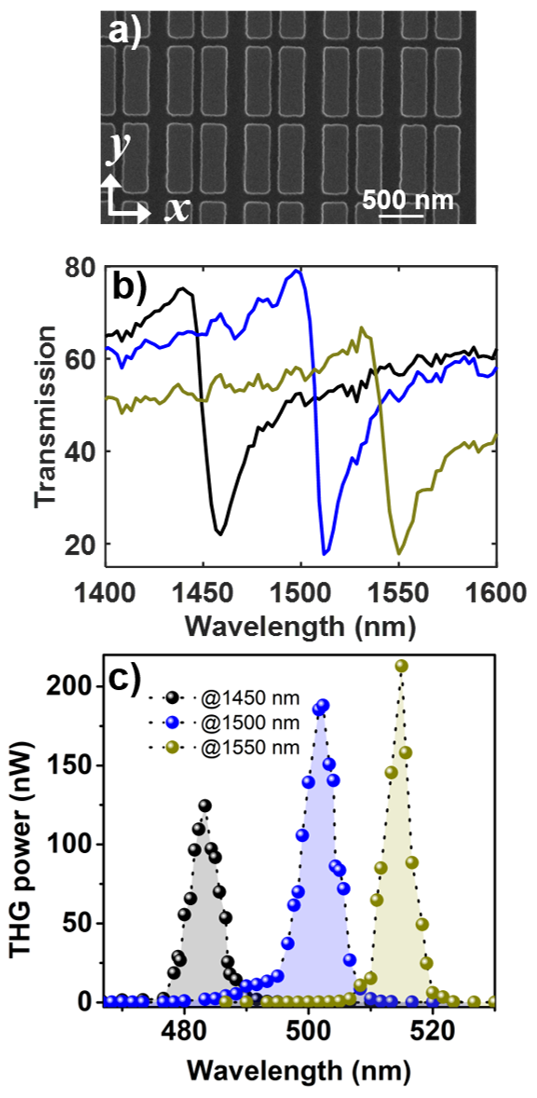}
\caption{(a) Scanning electron microscope images of the fabricated sample with designed resonance at 1500 nm. (b) Experimentally measured linear spectra. (c) The experimentally measured THG spectra of the samples.}
\label{fig:THG}
\end{figure}

The fabrication is done via the standard electro-lithography, similar to our previous work~\cite{xu2019dynamic}, where nanostructures are fabricated out of amorphous silicon on a glass substrate.
The scanning electron microscope (SEM) image of one fabricated metasurface sample with designed resonance at 1500~nm is shown in Fig.~\ref{fig:THG}(a). 
The measured dimensions of the fabricated metasurfaces $[w, L, x_0]$ are $[269, 805, 188]$~nm (resonant at 1450~nm), $[290, 765, 184]$~nm (resonant at 1500~nm), and $[310, 780, 184]$~nm (resonant at 1550~nm), respectively (see Fig.~S6 of the Supporting Information). While, the experimental and theoretical $x_0$ are identical, controlled by the lithography software, there are slight differences in the values of fabricated $w$ and $L$ from the theoretical targeting values as shown in Fig.~\ref{fig:inversedesign}(a)-\ref{fig:inversedesign}(c), due to fabrication imperfections. Such discrepancies, together with a possible minor inhomogeneity of the experimentally deposited silicon film, slightly affect the experimentally fabricated samples. It is worth noting that the lattice perturbations at the array's edge can also break the periodic boundary conditions leading to scattering of light in all directions into the free space, which weakens the total transmission or total reflection of the Fano resonance.

We first measured the linear transmission spectra of the three metasurfaces under plane wave normal incidence with the electric field polarized along the y-axis, as shown in Fig.~\ref{fig:THG}(b). Pronounced asymmetric Fano resonances are observed around the desired spectral positions. Subsequently, we perform the TH spectroscopy measurement. A femtosecond laser beam with 200~fs pulse width and 80 MHz repetition rate was focused by an aspheric lens with the focal length being 5 cm to a beam waist of 20~$\mathrm{\mu}$m. The pump polarization was adjusted along the $y$-axis, in order to excite the designed TD BIC state, and tuned ranging from 1400~nm to 1600~nm, with maximum mean power in the sample plane up to around 66~mW, leading to a maximum peak intensity value around $\mathrm{0.66~GW/cm^2}$. An objective with a numerical aperture (NA) NA=0.7 was used to collect the transmitted TH emission power in the forward direction (see Fig.~S7 in Section V. Experimental setup for nonlinear measurements of the Supporting Information). The experimentally measured TH signals from the three designed metasurfaces are shown in Fig.~\ref{fig:THG}(c). As can be seen, the TH signals are significantly enhanced around the resonances, while no THG enhancement is observed when the laser beam is polarized along $x$-axis (see Fig.~S8 in the Supporting Information). By comparison, we observed 400-fold enhancement of the TH signal at the resonance position under the y-polarized pump as compared to the case under $x$-polarized pump (Fig.~S8 in the Supporting Information). Similarly, by performing the nonlinear multipolar analysis, the TH signal is dominated by the TD excitation with small portions of MQ and EQ excitations, which exhibit the same  $C_2$ symmetry (Fig.~S9 in the Supporting Information). It further leads to a stronger TH emission in the first-order diffraction compared to the zero-order diffraction, due to the absence of coupling to these modes and the normal outgoing waves (see Fig.~S9 in the Supporting Information).

Another advantage of high-Q resonances is that the particular electric field distributions can also facilitate the optomechanical vibration process~\cite{Medeghini2018,Gan2019,Sun2016,Ivinskaya2018,Yi2018}. In other words, by designing a high-Q resonance through machine learning,  one can engineer the vibrational modes of nanoparticles based on the optomechanical processes, too. Here, by considering the designed metasurface shown in Fig.~\ref{fig:2}(d), we numerically investigate the optomechanical vibrations and mechanical mode excitations. The mechanical vibration is modelled via the following equation:

\begin{equation} \label{eq:OptoMech}
    \rho\frac{\partial^2u(r,t)}{\partial t^2}=\nabla \cdot \lbrace \left( 1+\beta\frac{\partial}{\partial t} \right) \left[ C: \frac{\left(\nabla u(r,t)\right)^{T}+\nabla u(r,t)}{2}  \right] \rbrace + F(r,t)
\end{equation}
where $u$ is the displacement field characterizing the mechanical vibration, the constant $\rho$, $C$, and $\beta$ represent mass density, stiffness tensor and decay time of the silicon material, respectively. $F(r,t)$ is the driving force induced by the electromagnetic field here.
Low-loss Si nanostructures can have significantly high laser damage threshold: $\thicksim \mathrm{400~GW/cm^2}$ or $\thicksim \mathrm{100~mJ/cm^2}$ at 250~fs light pump, and $\thicksim \mathrm{1000~GW/cm^2}$ or $\thicksim \mathrm{100~mJ/cm^2}$ at 100~fs light pump~\cite{shamonina2017world,Yang2015Nonlinear,makarov2015tuning}.
Here, the optical pump in our analysis is a single y-polarized laser pulse at the wavelength of 1500~nm with pulse duration 200 fs, peak intensity $\mathrm{I_0=50~GW/cm^2}$, at time $t_0=0$ s. Here we numerically calculate the effect of optical force on the mechanical vibration of our nanostructures using COMSOL Multiphysics. The optical force is determined by the time-averaged Maxwell stress tensor~\cite{jackson1999classical}:
\begin{equation} \label{eq:OptoMechTensor}
    T_{\alpha\beta}=\epsilon\left(E^{*}_{\alpha}E_{\beta}+ E_{\alpha}E^{*}_{\beta} - \delta_{\alpha\beta}|\mathbf{E}|^{2}\right) + \mu \left( H^{*}_{\alpha}H_{\beta} + H_{\alpha}H^{*}_{\beta} - \delta_{\alpha\beta}|\mathbf{H}|^{2}\right)
\end{equation}
where $\delta_{\alpha\beta}=1$ when $\alpha=\beta$ and $\delta_{\alpha\beta}=0$ otherwise.
The optical force in $\alpha$ direction is then derived from the induced optical near-field profiles using the following equation: 
\begin{equation} \label{eq:OptoMech2}
    F_{\alpha}=\int\left(\nabla\cdot T_{\alpha\beta}\right)_{\alpha}dV
\end{equation}

\begin{figure}[h!]
\centering
\includegraphics[width=0.85\textwidth]{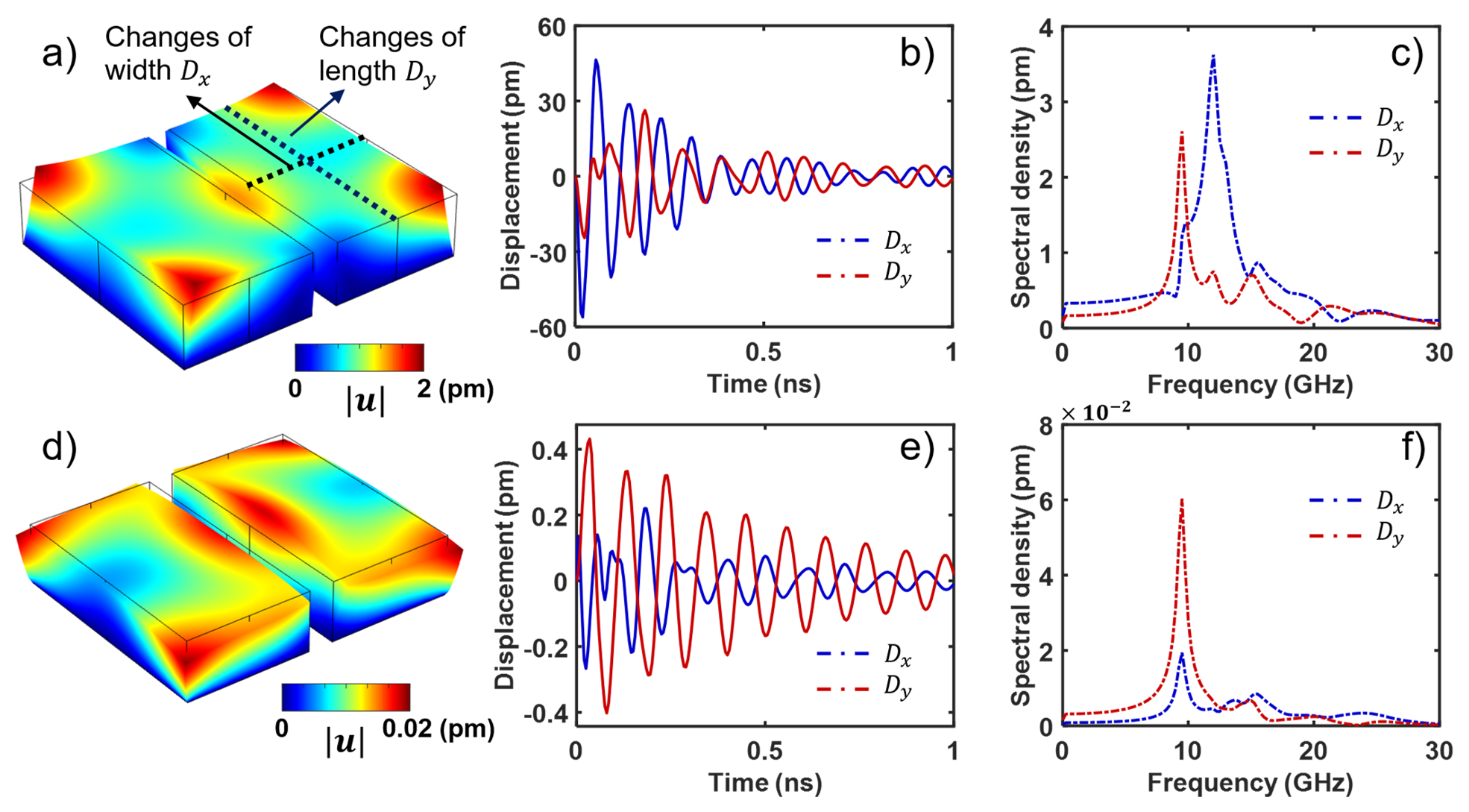}
\caption{(a-c) Optomechanic vibration under y-polarized pump: (a) Displacement of the nanobars after 1~ns. (b) The transient displacement $D_x$ and $D_y$. (c) Spectral densities of displacement $D_x$ and $D_y$. (d-f) Optomechanical vibration under $x$-polarized pump: (d) Displacement of the nanobars after 1~ns. (b) The transient displacement $D_x$ and $D_y$. (c) Spectral densities of displacement $D_x$ and $D_y$.}
\label{fig:OM}
\end{figure}

Due to the symmetric distributions of optical fields in these two nanobars shown in Fig.~\ref{fig:1}(d) and Fig.~\ref{fig:OM}(a), the corresponding vibration process for them are similar and have the same absolute values. Subsequently, we analyze the vibrations by measuring the changes in the width and length along the center of right nanobar in the unit cell (see Fig.~\ref{fig:OM}(a)). The vibration displacements along x and y directions, i.e. $D_x$ and $D_y$, are shown in Fig.~\ref{fig:OM}(b). 
To reveal the excitation of acoustic modes under optical pump, we further calculate the spectral density of the optomechanical vibrations. The corresponding spectral density is computed by the Fourier transfer analysis of this time-dependent vibrational amplitude and plotted in Fig.~\ref{fig:OM}(c) showing that the coherent phonon oscillation frequency for this nanobar is around  12~GHz for $D_x$ and 9.5~GHz for $D_y$. For comparison, Fig.~\ref{fig:OM}(d)-\ref{fig:OM}(f) shows the corresponding optomechanical vibrations when the pump is polarized along $x$-axis. The amplitude of optomechanical vibration is around 100-fold stronger from the metasurface under y-polarized pump incidence, which corresponds to the excitation of the designed TD BIC state. 
It is worth noting that the optomechanical response and the excitation of acoustic modes are dependent on the dimensions of the nanostructures. Therefore, By geometric tuning of the nanostructures, efficient excitation of different acoustic modes at other frequencies can be achieved.

\begin{figure}[h!]
\centering
\includegraphics[width=0.85\textwidth]{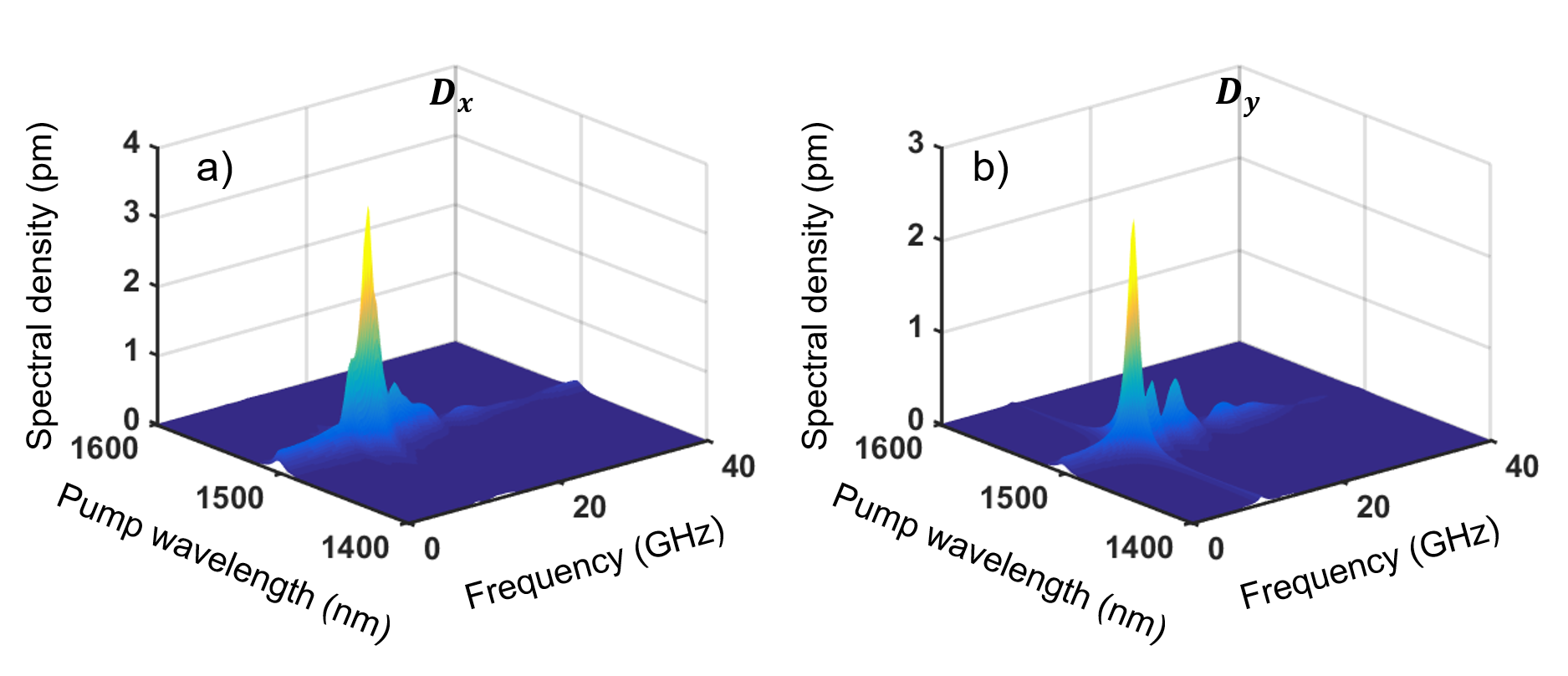}
\caption{The spectral density of D in the $x$ (a) and $y$ (b) directions for different laser pump wavelength.}
\label{fig:OM2}
\end{figure}

In order to demonstrate the importance of high-Q resonance on acoustic modes, we further calculate the spectral densities for different pump wavelengths under y-polarized pump incidence, as shown in the 3-dimensional map in Fig.~\ref{fig:OM2}. As can be seen, a significant peak with the frequency of 12~GHz for $D_x$ (9.5~GHz for $D_y$) appears for wavelength at 1500~nm, due to the excitation of TD BIC state at the optical pump. The excitation strength of acoustic mode decreases dramatically when the optical pump is away from the TD BIC state. We then estimate the feedback of the mechanical vibration on the resonant optical response. Based on the transient vibration shown in Fig.~\ref{fig:OM}(b), by assuming deform around 50-pm displacement in $x$ direction or 25-pm displacement in $y$ direction, the maximum sensitivity of the scattering response in such metasurfaces is around 0.055\% $\mathrm{pm^{-1}}$ (see Fig.~S10 in Supporting Information). Utilizing the high-quality TD BIC state, up to 4.5\% modulation of the transmission near the resonance can be expected when using a pulse laser with peak intensity $\mathrm{50~GW/cm^2}$ through the radiation force on the silicon nanostructures. These results suggest new opportunities for optomechanical applications such as light modulation and nanosensing with nanostructures.

\section{Conclusion}

To summarize, utilizing machine learning approach, we have demonstrated the inverse design of high-quality Fano resonant metasurfaces composed of two nanobars with scalable characteristics including the spectral position, line width and amplitude of the transmission. The Fano resonance is originated from the TD-BIC state, featuring a strong near-field enhancement and intense electric energy localized inside the nanobars. We further employ these metasurfaces to simultaneously enhance photon-photon and photon-phonon interactions and achieving 400+ folds THG enhancement and 100+ folds enhancement for optomechanical vibrations. Our proposed scalable metasurfaces suggest new opportunities to control and enhance light-matter interactions, showing promising applications for realizing optoacoustic nonlinear metasurfaces.


\acknowledgments 
We are grateful to Andrey Sukhorukov, Yue Sun and Camille Diffine for fruitful discussions. The authors acknowledge the funding support provided by the Australian Research Council (ARC). The work of A.E.M. was supported by UNSW Scientia Fellowship and ARC Discovery Project (DP170103778). M.R sincerely appreciates funding from ARC Discover Early Career Research Fellowship (DE170100250). D.S. acknowledges financial support from the Russian Foundation for Basic Research (Grants No. 18-02-00381, 19-02-00261) and the Australian Research Council (DE19010043). The authors appreciate the use of the Australian National Fabrication Facility (ANFF) – the ACT Node.   



\begin{thebibliography}{10}

\bibitem{Neshev2018Optical}
D.~Neshev and I.~Aharonovich, ``Optical metasurfaces: New generation building
  blocks for multi-functional optics,'' {\em Light-Sci. Appl.} {\bf 7}  (2018).

\bibitem{hail2019optical}
C.~U. Hail, A.-K.~U. Michel, D.~Poulikakos, {\em et~al.}, ``Optical
  metasurfaces: Evolving from passive to adaptive,'' {\em Advanced Optical
  Materials} {\bf 7}(14), 1801786  (2019).

\bibitem{chang2018optical}
S.~Chang, X.~Guo, and X.~Ni, ``Optical metasurfaces: Progress and
  applications,'' {\em Annu. Rev. Mater. Res.} {\bf 48}, 279--302  (2018).

\bibitem{Kuznetsov2016Optically}
A.~I. Kuznetsov, A.~E. Miroshnichenko, M.~L. Brongersma, {\em et~al.},
  ``Optically resonant dielectric nanostructures,'' {\em Science} {\bf 354},
  aag2472  (2016).

\bibitem{rahmani2017reversible}
M.~Rahmani, L.~Xu, A.~E. Miroshnichenko, {\em et~al.}, ``Reversible thermal
  tuning of all-dielectric metasurfaces,'' {\em Adv. Funct. Mater.} {\bf
  27}(31), 1700580  (2017).

\bibitem{miroshnichenko2010fano}
A.~E. Miroshnichenko, S.~Flach, and Y.~S. Kivshar, ``Fano resonances in
  nanoscale structures,'' {\em Reviews of Modern Physics} {\bf 82}(3), 2257
  (2010).

\bibitem{luk2010fano}
B.~Luk'yanchuk, N.~I. Zheludev, S.~A. Maier, {\em et~al.}, ``The fano resonance
  in plasmonic nanostructures and metamaterials,'' {\em Nat. Mater.} {\bf
  9}(9), 707  (2010).

\bibitem{khardikov2012giant}
V.~V. Khardikov, E.~O. Iarko, and S.~L. Prosvirnin, ``A giant red shift and
  enhancement of the light confinement in a planar array of dielectric bars,''
  {\em Journal of Optics} {\bf 14}(3), 035103  (2012).

\bibitem{rahmani2013fano}
M.~Rahmani, B.~Luk'yanchuk, and M.~Hong, ``Fano resonance in novel plasmonic
  nanostructures,'' {\em Laser \& Photonics Reviews} {\bf 7}(3), 329--349
  (2013).

\bibitem{gupta2016toroidal}
M.~Gupta and R.~Singh, ``Toroidal versus fano resonances in high q planar thz
  metamaterials,'' {\em Advanced Optical Materials} {\bf 4}(12), 2119--2125
  (2016).

\bibitem{Yang2015Nonlinear}
Y.~Yang, W.~Wang, A.~Boulesbaa, {\em et~al.}, ``Nonlinear fano-resonant
  dielectric metasurfaces,'' {\em Nano Lett.} {\bf 15}, 7388--7393  (2015).

\bibitem{tuz2018high}
V.~R. Tuz, V.~V. Khardikov, A.~S. Kupriianov, {\em et~al.}, ``High-quality
  trapped modes in all-dielectric metamaterials,'' {\em Optics Express} {\bf
  26}(3), 2905--2916  (2018).

\bibitem{campione2016broken}
S.~Campione, S.~Liu, L.~I. Basilio, {\em et~al.}, ``Broken symmetry dielectric
  resonators for high quality factor fano metasurfaces,'' {\em Acs Photonics}
  {\bf 3}(12), 2362--2367  (2016).

\bibitem{zangeneh2019reversible}
K.~Zangeneh~Kamali, L.~Xu, J.~Ward, {\em et~al.}, ``Reversible image contrast
  manipulation with thermally tunable dielectric metasurfaces,'' {\em Small}
  {\bf 15}(15), 1805142  (2019).

\bibitem{xu2019dynamic}
L.~Xu, K.~Zangeneh~Kamali, L.~Huang, {\em et~al.}, ``Dynamic nonlinear image
  tuning through magnetic dipole quasi-bic ultrathin resonators,'' {\em Adv.
  Sci.} {\bf 6}, 1802119  (2019).

\bibitem{jain2015electric}
A.~Jain, P.~Moitra, T.~Koschny, {\em et~al.}, ``Electric and magnetic response
  in dielectric dark states for low loss subwavelength optical meta atoms,''
  {\em Advanced Optical Materials} {\bf 3}(10), 1431--1438  (2015).

\bibitem{hao2009tunability}
F.~Hao, P.~Nordlander, Y.~Sonnefraud, {\em et~al.}, ``Tunability of subradiant
  dipolar and fano-type plasmon resonances in metallic ring/disk cavities:
  implications for nanoscale optical sensing,'' {\em ACS nano} {\bf 3}(3),
  643--652  (2009).

\bibitem{liu2011three}
N.~Liu, M.~Hentschel, T.~Weiss, {\em et~al.}, ``Three-dimensional plasmon
  rulers,'' {\em Science} {\bf 332}(6036), 1407--1410  (2011).

\bibitem{hao2008symmetry}
F.~Hao, Y.~Sonnefraud, P.~V. Dorpe, {\em et~al.}, ``Symmetry breaking in
  plasmonic nanocavities: subradiant lspr sensing and a tunable fano
  resonance,'' {\em Nano letters} {\bf 8}(11), 3983--3988  (2008).

\bibitem{bao2010fano}
K.~Bao, N.~A. Mirin, and P.~Nordlander, ``Fano resonances in planar silver
  nanosphere clusters,'' {\em Applied Physics A} {\bf 100}(2), 333--339
  (2010).

\bibitem{halas2011plasmons}
N.~J. Halas, S.~Lal, W.-S. Chang, {\em et~al.}, ``Plasmons in strongly coupled
  metallic nanostructures,'' {\em Chemical reviews} {\bf 111}(6), 3913--3961
  (2011).

\bibitem{hentschel2010transition}
M.~Hentschel, M.~Saliba, R.~Vogelgesang, {\em et~al.}, ``Transition from
  isolated to collective modes in plasmonic oligomers,'' {\em Nano letters}
  {\bf 10}(7), 2721--2726  (2010).

\bibitem{rahmani2012subgroup}
M.~Rahmani, D.~Y. Lei, V.~Giannini, {\em et~al.}, ``Subgroup decomposition of
  plasmonic resonances in hybrid oligomers: modeling the resonance lineshape,''
  {\em Nano letters} {\bf 12}(4), 2101--2106  (2012).

\bibitem{hsu2016bound}
C.~W. Hsu, B.~Zhen, A.~D. Stone, {\em et~al.}, ``Bound states in the
  continuum,'' {\em Nature Reviews Materials} {\bf 1}(9), 1--13  (2016).

\bibitem{carletti2018giant}
L.~Carletti, K.~Koshelev, C.~De~Angelis, {\em et~al.}, ``Giant nonlinear
  response at the nanoscale driven by bound states in the continuum,'' {\em
  Physical Review Letters} {\bf 121}(3), 033903  (2018).

\bibitem{koshelev2018asymmetric}
K.~Koshelev, S.~Lepeshov, M.~Liu, {\em et~al.}, ``Asymmetric metasurfaces with
  high-q resonances governed by bound states in the continuum,'' {\em Physical
  review letters} {\bf 121}(19), 193903  (2018).

\bibitem{liu2018extreme}
M.~Liu and D.-Y. Choi, ``Extreme huygensâ€™ metasurfaces based on quasi-bound
  states in the continuum,'' {\em Nano letters} {\bf 18}(12), 8062--8069
  (2018).

\bibitem{he2018toroidal}
Y.~He, G.~Guo, T.~Feng, {\em et~al.}, ``Toroidal dipole bound states in the
  continuum,'' {\em Physical Review B} {\bf 98}(16), 161112  (2018).

\bibitem{Koshelev2019}
K.~Koshelev, Y.~Tang, K.~Li, {\em et~al.}, ``{Nonlinear Metasurfaces Governed
  by Bound States in the Continuum},'' {\em ACS Photonics}   (2019).

\bibitem{luo2020excitation}
X.~Luo, X.~Li, T.~Lang, {\em et~al.}, ``Excitation of high q toroidal dipole
  resonance in an all-dielectric metasurface,'' {\em Optical Materials Express}
  {\bf 10}(2), 358--368  (2020).

\bibitem{carletti2019high}
L.~Carletti, S.~S. Kruk, A.~A. Bogdanov, {\em et~al.}, ``High-harmonic
  generation at the nanoscale boosted by bound states in the continuum,'' {\em
  Physical Review Research} {\bf 1}(2), 023016  (2019).

\bibitem{gandhi2020gain}
H.~K. Gandhi, D.~Rocco, L.~Carletti, {\em et~al.}, ``Gain-loss engineering of
  bound states in the continuum for enhanced nonlinear response in dielectric
  nanocavities,'' {\em Optics Express} {\bf 28}(3), 3009--3016  (2020).

\bibitem{ndangali2010electromagnetic}
R.~F. Ndangali and S.~V. Shabanov, ``Electromagnetic bound states in the
  radiation continuum for periodic double arrays of subwavelength dielectric
  cylinders,'' {\em Journal of mathematical physics} {\bf 51}(10), 102901
  (2010).

\bibitem{Gili2016Monolithic}
V.~F. Gili, L.~Carletti, A.~Locatelli, {\em et~al.}, ``Monolithic {AlGaAs}
  second-harmonic nanoantennas,'' {\em Opt. Express} {\bf 24}, 15965  (2016).

\bibitem{Liu2016Resonantly}
S.~Liu, M.~B. Sinclair, S.~Saravi, {\em et~al.}, ``Resonantly enhanced
  second-harmonic generation using {III}{\textendash}{V} semiconductor
  all-dielectric metasurfaces,'' {\em Nano Lett.} {\bf 16}, 5426--5432  (2016).

\bibitem{Camacho-Morales2016Nonlinear}
R.~Camacho-Morales, M.~Rahmani, S.~Kruk, {\em et~al.}, ``Nonlinear generation
  of vector beams from {AlGaAs} nanoantennas,'' {\em Nano Lett.} {\bf 16},
  7191--7197  (2016).

\bibitem{Grinblat2016Efficient}
G.~Grinblat, Y.~Li, M.~P. Nielsen, {\em et~al.}, ``Efficient third harmonic
  generation and nonlinear subwavelength imaging at a higher-order anapole mode
  in a single germanium nanodisk,'' {\em {ACS} Nano} {\bf 11}, 953--960
  (2016).

\bibitem{Grinblat2016Enhanced}
G.~Grinblat, Y.~Li, M.~P. Nielsen, {\em et~al.}, ``Enhanced third harmonic
  generation in single germanium nanodisks excited at the anapole mode,'' {\em
  Nano Lett.} {\bf 16}, 4635--4640  (2016).

\bibitem{shcherbakov2015ultrafast}
M.~R. Shcherbakov, P.~P. Vabishchevich, A.~S. Shorokhov, {\em et~al.},
  ``Ultrafast all-optical switching with magnetic resonances in nonlinear
  dielectric nanostructures,'' {\em Nano Lett.} {\bf 15}(10), 6985--6990
  (2015).

\bibitem{Melik-Gaykazyan2017Third}
E.~V. Melik-Gaykazyan, M.~R. Shcherbakov, A.~S. Shorokhov, {\em et~al.},
  ``Third-harmonic generation from mie-type resonances of isolated
  all-dielectric nanoparticles,'' {\em Philos. Trans. Royal Soc. A} {\bf 375},
  20160281  (2017).

\bibitem{melik2017selective}
E.~V. Melik-Gaykazyan, S.~S. Kruk, R.~Camacho-Morales, {\em et~al.},
  ``Selective third-harmonic generation by structured light in mie-resonant
  nanoparticles,'' {\em ACS Photonics} {\bf 5}(3), 728--733  (2017).

\bibitem{miroshnichenko2015nonradiating}
A.~E. Miroshnichenko, A.~B. Evlyukhin, Y.~F. Yu, {\em et~al.}, ``Nonradiating
  anapole modes in dielectric nanoparticles,'' {\em Nat. Commun.} {\bf 6}
  (2015).

\bibitem{Xu2018Boosting}
L.~Xu, M.~Rahmani, K.~Z. Kamali, {\em et~al.}, ``Boosting third-harmonic
  generation by a mirror-enhanced anapole resonator,'' {\em Light: Sci. Appl.}
  {\bf 7}, 44  (2018).

\bibitem{Carletti2015Enhanced}
L.~Carletti, A.~Locatelli, O.~Stepanenko, {\em et~al.}, ``Enhanced
  second-harmonic generation from magnetic resonance in {AlGaAs}
  nanoantennas,'' {\em Opt. Express} {\bf 23}, 26544  (2015).

\bibitem{molesky2018inverse}
S.~Molesky, Z.~Lin, A.~Y. Piggott, {\em et~al.}, ``Inverse design in
  nanophotonics,'' {\em Nature Photonics} {\bf 12}(11), 659--670  (2018).

\bibitem{liu2018training}
D.~Liu, Y.~Tan, E.~Khoram, {\em et~al.}, ``Training deep neural networks for
  the inverse design of nanophotonic structures,'' {\em ACS Photonics} {\bf
  5}(4), 1365--1369  (2018).

\bibitem{mirzaei2014superscattering}
A.~Mirzaei, A.~E. Miroshnichenko, I.~V. Shadrivov, {\em et~al.},
  ``Superscattering of light optimized by a genetic algorithm,'' {\em Applied
  Physics Letters} {\bf 105}(1), 011109  (2014).

\bibitem{nadell2019deep}
C.~C. Nadell, B.~Huang, J.~M. Malof, {\em et~al.}, ``Deep learning for
  accelerated all-dielectric metasurface design,'' {\em Optics Express} {\bf
  27}(20), 27523--27535  (2019).

\bibitem{kiarashinejad2020deep}
Y.~Kiarashinejad, S.~Abdollahramezani, and A.~Adibi, ``Deep learning approach
  based on dimensionality reduction for designing electromagnetic
  nanostructures,'' {\em npj Computational Materials} {\bf 6}(1), 1--12
  (2020).

\bibitem{ma2018deep}
W.~Ma, F.~Cheng, and Y.~Liu, ``Deep-learning-enabled on-demand design of chiral
  metamaterials,'' {\em ACS nano} {\bf 12}(6), 6326--6334  (2018).

\bibitem{yao2019intelligent}
K.~Yao, R.~Unni, and Y.~Zheng, ``Intelligent nanophotonics: merging photonics
  and artificial intelligence at the nanoscale,'' {\em Nanophotonics} {\bf
  8}(3), 339--366  (2019).

\bibitem{liu2019integrated}
Z.~Liu, X.~Liu, Z.~Xiao, {\em et~al.}, ``Integrated nanophotonic wavelength
  router based on an intelligent algorithm,'' {\em Optica} {\bf 6}(10),
  1367--1373  (2019).

\bibitem{gao2019bidirectional}
L.~Gao, X.~Li, D.~Liu, {\em et~al.}, ``A bidirectional deep neural network for
  accurate silicon color design,'' {\em Advanced Materials} {\bf 31}(51),
  1905467  (2019).

\bibitem{zhang2019artificial}
Q.~Zhang, H.~Yu, M.~Barbiero, {\em et~al.}, ``Artificial neural networks
  enabled by nanophotonics,'' {\em Light: Science \& Applications} {\bf 8}(1),
  1--14  (2019).

\bibitem{jiang2019simulator}
J.~Jiang and J.~A. Fan, ``Simulator-based training of generative neural
  networks for the inverse design of metasurfaces,'' {\em Nanophotonics}
  (2019).

\bibitem{so2019simultaneous}
S.~So, J.~Mun, and J.~Rho, ``Simultaneous inverse design of materials and
  structures via deep learning: Demonstration of dipole resonance engineering
  using core--shell nanoparticles,'' {\em ACS applied materials \& interfaces}
  {\bf 11}(27), 24264--24268  (2019).

\bibitem{jiang2019global}
J.~Jiang and J.~A. Fan, ``Global optimization of dielectric metasurfaces using
  a physics-driven neural network,'' {\em Nano letters} {\bf 19}(8), 5366--5372
   (2019).

\bibitem{wiecha2019pushing}
P.~R. Wiecha, A.~Lecestre, N.~Mallet, {\em et~al.}, ``Pushing the limits of
  optical information storage using deep learning,'' {\em Nature
  nanotechnology} {\bf 14}(3), 237--244  (2019).

\bibitem{piggott2015inverse}
A.~Y. Piggott, J.~Lu, K.~G. Lagoudakis, {\em et~al.}, ``Inverse design and
  demonstration of a compact and broadband on-chip wavelength demultiplexer,''
  {\em Nature Photonics} {\bf 9}(6), 374--377  (2015).

\bibitem{asano2018optimization}
T.~Asano and S.~Noda, ``Optimization of photonic crystal nanocavities based on
  deep learning,'' {\em Optics Express} {\bf 26}(25), 32704--32717  (2018).

\bibitem{carleo2019machine}
G.~Carleo, I.~Cirac, K.~Cranmer, {\em et~al.}, ``Machine learning and the
  physical sciences,'' {\em Reviews of Modern Physics} {\bf 91}(4), 045002
  (2019).

\bibitem{carleo2017solving}
G.~Carleo and M.~Troyer, ``Solving the quantum many-body problem with
  artificial neural networks,'' {\em Science} {\bf 355}(6325), 602--606
  (2017).

\bibitem{hertz2018introduction}
J.~A. Hertz, {\em Introduction to the theory of neural computation}, CRC Press
  (2018).

\bibitem{hornik1989multilayer}
K.~Hornik, M.~Stinchcombe, H.~White, {\em et~al.}, ``Multilayer feedforward
  networks are universal approximators.,'' {\em Neural networks} {\bf 2}(5),
  359--366  (1989).

\bibitem{Aouani2014Third}
H.~Aouani, M.~Rahmani, M.~Navarro-C{\'{\i}}a, {\em et~al.},
  ``Third-harmonic-upconversion enhancement from a single semiconductor
  nanoparticle coupled to a plasmonic antenna,'' {\em Nat. Nanotechnol.} {\bf
  9}, 290--294  (2014).

\bibitem{lee2014giant}
J.~Lee, M.~Tymchenko, C.~Argyropoulos, {\em et~al.}, ``Giant nonlinear response
  from plasmonic metasurfaces coupled to intersubband transitions,'' {\em
  Nature} {\bf 511}(7507), 65--69  (2014).

\bibitem{merklein2015enhancing}
M.~Merklein, I.~V. Kabakova, T.~F. B{\"u}ttner, {\em et~al.}, ``Enhancing and
  inhibiting stimulated brillouin scattering in photonic integrated circuits,''
  {\em Nature communications} {\bf 6}(1), 1--8  (2015).

\bibitem{pant2011chip}
R.~Pant, C.~G. Poulton, D.-Y. Choi, {\em et~al.}, ``On-chip stimulated
  brillouin scattering,'' {\em Optics Express} {\bf 19}(9), 8285--8290  (2011).

\bibitem{VanLaer2015}
R.~{Van Laer}, B.~Kuyken, D.~{Van Thourhout}, {\em et~al.}, ``{Interaction
  between light and highly confined hypersound in a silicon photonic
  nanowire},'' {\em Nature Photonics}   (2015).

\bibitem{Grinblat2014}
G.~Grinblat, M.~Rahmani, E.~Cort{\'{e}}s, {\em et~al.}, ``{High-efficiency
  second harmonic generation from a single hybrid Zno nanowire/au plasmonic
  nano-oligomer},'' {\em Nano Letters}   (2014).

\bibitem{DellaPicca2016}
F.~{Della Picca}, R.~Berte, M.~Rahmani, {\em et~al.}, ``{Tailored Hypersound
  Generation in Single Plasmonic Nanoantennas},'' {\em Nano Letters}   (2016).

\bibitem{Aouani2015}
H.~Aouani, M.~Navarro-C{\'{i}}a, M.~Rahmani, {\em et~al.}, ``{Unveiling the
  Origin of Third Harmonic Generation in Hybrid ITO-Plasmonic Crystals},'' {\em
  Advanced Optical Materials}   (2015).

\bibitem{Chen2011}
Y.~S. Chen, W.~Frey, S.~Kim, {\em et~al.}, ``{Silica-coated gold nanorods as
  photoacoustic signal nanoamplifiers},'' {\em Nano Letters}   (2011).

\bibitem{Moreaux2001}
L.~Moreaux, O.~Sandre, S.~Charpak, {\em et~al.}, ``{Coherent scattering in
  multi-harmonic light microscopy},'' {\em Biophysical Journal}   (2001).

\bibitem{Chollet2015}
F.~Chollet, ``{Keras: Deep Learning library for Theano and TensorFlow},'' {\em
  GitHub Repository}   (2015).

\bibitem{Moharam1995}
M.~G. Moharam, T.~K. Gaylord, E.~B. Grann, {\em et~al.}, ``{Formulation for
  stable and efficient implementation of the rigorous coupled-wave analysis of
  binary gratings},'' {\em Journal of the Optical Society of America A}
  (1995).

\bibitem{Hugonin2005rcwa}
J.~P. Hugonin and P.~Lalanne, ``Reticolo software for grating analysis,'' {\em
  Institute of Optics Graduates School, Orsay, France}   (2005).

\bibitem{Arbabi2015}
A.~Arbabi, Y.~Horie, M.~Bagheri, {\em et~al.}, ``{Dielectric metasurfaces for
  complete control of phase and polarization with subwavelength spatial
  resolution and high transmission},'' {\em Nature Nanotechnology}   (2015).

\bibitem{Peurifoy2018}
J.~Peurifoy, Y.~Shen, L.~Jing, {\em et~al.}, ``{Nanophotonic particle
  simulation and inverse design using artificial neural networks},'' {\em
  Science Advances}   (2018).

\bibitem{Maksymov2011}
I.~S. Maksymov and A.~E. Miroshnichenko, ``{Active control over nanofocusing
  with nanorod plasmonic antennas},'' {\em Optics Express}   (2011).

\bibitem{Galli2009}
M.~Galli, S.~L. Portalupi, M.~Belotti, {\em et~al.}, ``{Light scattering and
  Fano resonances in high-Q photonic crystal nanocavities},'' {\em Applied
  Physics Letters}   (2009).

\bibitem{Medeghini2018}
F.~Medeghini, A.~Crut, M.~Gandolfi, {\em et~al.}, ``{Controlling the Quality
  Factor of a Single Acoustic Nanoresonator by Tuning its Morphology},'' {\em
  Nano Letters}   (2018).

\bibitem{Gan2019}
Y.~Gan and Z.~Sun, ``{Crystal structure dependence of the breathing vibration
  of individual gold nanodisks induced by the ultrafast laser},'' {\em Applied
  Optics}   (2019).

\bibitem{Sun2016}
Y.~Sun, S.~V. Suchkov, A.~E. Miroshnichenko, {\em et~al.}, ``{Opto-mechanical
  interactions in nanoparticles with magnetic light},'' in {\em Optics InfoBase
  Conference Papers},   (2016).

\bibitem{Ivinskaya2018}
A.~Ivinskaya, N.~Kostina, A.~Proskurin, {\em et~al.}, ``{Optomechanical
  Manipulation with Hyperbolic Metasurfaces},'' {\em ACS Photonics}   (2018).

\bibitem{Yi2018}
C.~Yi, M.~N. Su, P.~D. Dongare, {\em et~al.}, ``{Polycrystallinity of
  Lithographically Fabricated Plasmonic Nanostructures Dominates Their Acoustic
  Vibrational Damping},'' {\em Nano Letters}   (2018).

\bibitem{shamonina2017world}
E.~Shamonina, {\em World Scientific Handbook of Metamaterials and Plasmonics},
  World Scientific, Oxford  (2017).

\bibitem{makarov2015tuning}
S.~Makarov, S.~Kudryashov, I.~Mukhin, {\em et~al.}, ``Tuning of magnetic
  optical response in a dielectric nanoparticle by ultrafast photoexcitation of
  dense electron--hole plasma,'' {\em Nano letters} {\bf 15}(9), 6187--6192
  (2015).

\bibitem{jackson1999classical}
J.~D. Jackson, {\em Classical Electrodynamics}, New York: John Wiley \& Sons
  (1999).

\end{thebibliography}





\end{spacing}
\end{document}